\documentclass[%
reprint,
superscriptaddress,
amsmath,amssymb,
aps,
prl,
longbibliography,
]{revtex4-1}

\usepackage{pdfpages} 
\usepackage{dirtytalk}
\usepackage{mathtools}
\usepackage{pdfpages}
\usepackage{mhchem}
\usepackage{graphicx}
\usepackage{dcolumn}
\usepackage{bm}
\usepackage{hyperref}
\hypersetup{
colorlinks = true,   
urlcolor = blue,    
linkcolor = magenta,   
citecolor = blue  
}
\usepackage{float}
\usepackage{color}

\makeatletter
\AtBeginDocument{\let\LS@rot\@undefined}
\makeatother

\begin{document}
\preprint{APS/123-QED}
\title{First- and Second-Order Topological Superconductivity and Temperature-Driven
\\ \hspace{-6mm}Topological
Phase Transitions in the Extended Hubbard Model with Spin-Orbit Coupling}
\author{Majid Kheirkhah}
\email{kheirkhah@ualberta.ca}
\affiliation{Department of Physics, University of Alberta, Edmonton, Alberta T6G 2E1, Canada }

\author{Zhongbo Yan}
\affiliation{School of Physics, Sun Yat-Sen University, Guangzhou 510275, China}

\author{Yuki Nagai}
\affiliation{CCSE, Japan Atomic Energy Agency, 178-4-4, Wakashiba, Kashiwa, Chiba, 277-0871, Japan} 
\affiliation{Mathematical Science Team, RIKEN Center for Advanced Intelligence Project (AIP), 1-4-1 Nihonbashi, Chuo-ku, Tokyo 103-0027, Japan}

\author{Frank Marsiglio}
\affiliation{Department of Physics, University of Alberta, Edmonton, Alberta T6G 2E1, Canada }
\date{\today}
\begin{abstract}  
The combination of spin-orbit coupling with interactions results in many exotic phases of matter. In this Letter, we investigate the superconducting pairing instability of the
two-dimensional extended Hubbard model with both Rashba and Dresselhaus spin-orbit coupling within the mean-field level at both zero and finite temperature. We find that both first- and second-order time-reversal symmetry breaking topological gapped phases can be achieved under appropriate parameters and temperature regimes due to the presence of a favored even-parity $s+id$-wave pairing even in the absence of an external magnetic field or intrinsic magnetism. This results in two branches of chiral Majorana edge states on each edge or a single zero-energy Majorana corner state at each corner of the sample. Interestingly, we also find that not only does tuning the doping level lead to a direct topological phase transition between these two distinct topological gapped phases, but also using the temperature as a highly controllable and reversible tuning knob leads to different direct temperature-driven topological phase transitions between gapped and gapless topological superconducting phases. Our findings suggest new possibilities in interacting spin-orbit coupled systems by unifying both first- and higher-order topological superconductors in a simple but realistic microscopic model.

\end{abstract}

\pacs{Valid PACS appear here}
\maketitle
{\it Introduction.---}Spin-orbit coupling (SOC) is ubiquitous in condensed matter systems and responsible for many remarkable phenomena \cite{rashba1959symmetry, rashba2015symmetry, PhysRev.96.280, PhysRev.95.568, PhysRev.100.580, PhysRev.98.368, ganichev2014interplay, manchon2015new, bihlmayer2015focus, PhysRevLett.87.037004, PhysRevLett.92.097001,PhysRevLett.75.2004, PhysRevLett.121.157003,PhysRevB.93.220505, PhysRevB.95.064509, Galitski2013}. 
In recent years, a surge of research interest in  
SOC was stimulated by the discovery that SOC plays a critical 
role in realizing various topological
phases, ranging from noninteracting or weakly correlated topological
insulators (TIs) and topological superconductors (TSCs) to strongly correlated
topological phases \cite{moore2010birth,Pesin2010,qi2011topological,hasan2010colloquium}. Among them, TSCs are noticeable as they harbor  Majorana modes which are believed to be a possibility for the building blocks of 
topological quantum computation \cite{PhysRevLett.86.268,  kitaev2003fault, RevModPhys.80.1083, alicea2012new, lian2018topological}. While odd-parity superconductors 
generally provide a natural realization of TSCs \cite{sato2010topological, PhysRevLett.105.097001, volovik1999fermion, read2000paired,kitaev2001unpaired, yonezawa2017thermodynamic}, their scarcity in nature 
turns out to be a serious obstacle from an experimental point of view. 
Fortunately, SOC enables the realization of effective odd-parity superconductivity (SC)
on the basis of abundant even-parity SC, providing
a more readily accessible route for the realization of TSCs \cite{fu2008superconducting,oreg2010helical,Lutchyn2010, PhysRevLett.103.020401, PhysRevLett.104.040502, PhysRevB.81.125318,wong2012majorana,zhang2013time}. Over the 
past decade, remarkable progress along this route has been witnessed \cite{mourik2012signatures,das2012zero,rokhinson2012fractional,deng2012anomalous,nadj2014observation,Zhang2018quantized,Sun2016Majorana,wang2018evidence,Lutchyn2018}.

Very recently, a new class of topological phases, named higher-order TIs and TSCs, have emerged and attracted 
a great deal of attention because of the enrichment of boundary physics and the occurrence of 
new possibilities for topological phase transitions \cite{benalcazar2017quantized,benalcazar2017electric, schindler2018higher, PhysRevLett.122.236401, langbehn2017reflection, ezawa2018higher, khalaf2018higher,geier2018second, franca2018anomalous,   cualuguaru2019higher, schindler2018higher_bis,imhof2018topolectrical,serra2018observation,peterson2018quantized,zhang2019second,Niu2020transition}. The word ``order'' in this context gives the codimension of the gapless boundary modes, namely, an $n$th order TI or TSC has gapless boundary modes with codimension $n$. As the gapless boundary modes of all conventional TIs and TSCs have $n=1$, they thus belong to the first-order topological phases
in this language.

Because higher-order TSCs provide new platforms of Majorana modes, their potential 
application in topological quantum computation has triggered quite a few theoretical proposals 
on their experimental realizations \cite{zhu2018tunable,yan2018majorana, wang2018weak,wang2018high, wu2019higher,hsu2018majorana,  liu2018majorana, volpez2019second, Zhang2019hinge, Yang2019hinge, PhysRevLett.123.060402,PhysRevLett.123.177001,PhysRevB.100.205406, PhysRevB.100.020509,PhysRevResearch.2.012018,Pan2018SOTSC, PhysRevResearch.1.032017, PhysRevB.100.075415,Zhang2019hoscb, ahn2019higher, PhysRevB.101.104502,ghorashi2019vortex,Wu2019swave,Wu2019hoscb,Hsu2019HOSC}. However, the superconducting pairings in previous 
works were mostly introduced phenomenologically, 
and realistic microscopic models for higher-order TSCs are still generally lacking. Over the past decade, the Hubbard model with SOC and on site interaction, 
as one of the simplest microscopic models for first-order TSCs, has been extensively studied in both condensed matter and in the cold atom communities
\cite{Scheurer2015,Greco2018,Zhang2008swave,Gong2012Majorana,Zhu2011Majorana,Liu2012Majorana,yan2015topological}. 
In this Letter, we extend the Hubbard model in two dimensions to include both on site (repulsive) and inter site (attractive) interactions and investigate its even-parity superconducting pairing instability at the mean-field level \cite{RevModPhys.62.113}.

Our study reveals that depending on the temperature and the parameters of the model, the leading pairing channel can be $d$-, $s$-, or $s+id$-wave \cite{tsuchiura1995stability, hutchinson2019superconducting}. Remarkably, we find when the $s+id$-wave is favored, a first-order TSC with two branches of chiral edge states and a second-order TSC with four Majorana corner modes, as well as a direct topological phase transition between them, can be realized by tuning the Fermi surface (FS) structure, even in the absence of a magnetic field or magnetism. Furthermore, we show that the temperature itself is a highly controllable and reversible tuning knob to drive topological phase transitions in this system.

{\it Theoretical formalism.---}The two-dimensional extended Hubbard model, which provides a simple description for short-ranged interacting systems \cite{RevModPhys.62.113}, reads as
\begin{align}
 H=& -t \sum_{\langle i,j \rangle, \alpha} c^{\dagger}_{i, \alpha} c_{j, \alpha}
+ \text{H.c.} - \mu \sum_{i,\alpha} c^{\dagger}_{i, \alpha}  c_{i,\alpha}  \nonumber
\\&
+i\lambda_R \sum_{i, \alpha,\beta} 
(c^{\dagger}_{i, \alpha} s_{y}^{\alpha \beta}  c_{i+\hat{x}, \beta}
-c^{\dagger}_{i, \alpha} s_{x}^{\alpha \beta}  c_{i+\hat{y}, \beta}) + \text{H.c.} \nonumber
\\&
+i\lambda_D \sum_{i, \alpha,\beta} 
(c^{\dagger}_{i, \alpha} s_{y}^{\alpha \beta}  c_{i+\hat{y}, \beta}
-c^{\dagger}_{i, \alpha} s_{x}^{\alpha \beta}  c_{i+\hat{x}, \beta}) + \text{H.c.} \nonumber
\\&
+U \sum_{i}  \hat{n}_{i, \uparrow} \hat{n}_{i, \downarrow} + \dfrac{V}{2}\sum_{i, j, \alpha}  \hat{n}_{i,\alpha} \hat{n}_{j,\Bar{\alpha}},
\end{align}
where $\langle i,j\rangle$ denotes summation over nearest-neighbor sites, $c^{\dagger}_{i,\alpha} (c_{i,\alpha})$ is the creation (annihilation) operator at site $i$ with spin $\alpha =(\uparrow, \downarrow)$ , $\hat{n}_{i,\alpha} = c^{\dagger}_{i, \alpha} c_{i, \alpha}$, $t$ is the nearest-neighbor hopping amplitude, $\mu$ is the chemical potential, $\lambda_R$ ($\lambda_D$) is the Rashba (Dresselhaus) SOC amplitude, $U$ is the on site repulsive ($U > 0$) interaction strength, and $V$ is the nearest-neighbor attractive ($V<0$) interaction strength. The unit vector along the $x$ ($y$) direction is represented by $\hat{x}$ ($\hat{y}$), and $s_{x,y}$ are Pauli matrices in spin space. The abbreviation \text{H.c.} stands for Hermitian conjugation, and the symbol $i$ in the beginning of both second and third lines (and elsewhere) is taken to denote the pure imaginary number and should not be confused with the site index which generally occurs as subscripts.

Although in the presence of SOC, odd- and even-parity pairings can generally coexist, we restrict ourselves to even-parity pairing for the sake of clarity and simplicity. Accordingly, the Bogoliubov-de Gennes (BdG) Hamiltonian at the mean-field level in momentum space (see Supplemental Material \cite{supplemental})  can be rewritten as $H=\frac{1}{2}\sum_{\bm{k}} \Psi^{\dagger}_{\bm{k}}\mathcal{H}(\bm{k})\Psi_{\bm{k}}$, with
$\Psi^{\text{T}}_{\bm{k}}=
(c_{\bm{k} \uparrow},~ c_{\bm{k} \downarrow},~c^{\dagger}_{-\bm{k} \downarrow},-c^{\dagger}_{-\bm{k} \uparrow})$ and 
\begin{align}
\hspace{-2mm}
\mathcal{H}(\bm{k}) = \tau_z \big\{ \xi(\bm{k}) s_0
+ l_x(\bm{k}) s_x + l_y(\bm{k}) s_y \big\}
+ \Delta(\bm{k}) \tau_x s_0,
\label{Hami_1}
\end{align}
where $\tau_{x,y,z}$ are Pauli matrices in particle-hole 
space, and $\xi_{\bm{k}} = -2t(\cos k_x+\cos k_y) -\mu$ is the kinetic 
energy measured from the Fermi energy;  
$\bm{l}(\bm{k}) = (l_x(\bm{k}), l_y(\bm{k}))$ is the SOC vector, 
with $\bm{l}(\bm{k}) = \bm{l}_R(\bm{k}) + \bm{l}_D(\bm{k})$, where
$\bm{l}_R(\bm{k}) = 2\lambda_R(\sin k_y, -\sin k_x)$ 
and $\bm{l}_D(\bm{k}) = 2\lambda_D(\sin k_x, -\sin k_y)$
represents the Rashba and Dresselhaus SOC, respectively. The superconducting order parameter is given by
\begin{equation}
\Delta(\bm{k}) = \Delta_{0}^{c} + \Delta_{s}^{c}\eta_{s}({\bm{k}}) + \Delta_{d}^{c}\eta_{d}({\bm{k}}),
\end{equation}
where $\Delta_{0}^{c}$, $\Delta_{s}^{c}$, and $\Delta_{d}^{c}$ are  momentum-independent complex numbers that represent on site $s$-, extended $s$-, and $d$-wave SC, respectively. The three pairing amplitudes satisfy the following self-consistent superconducting gap equations, 
\begin{align}
\Delta_{0}^{c} &= -\dfrac{U}{4N} \sum_{\bm{k}, \sigma} \Delta(\bm{k})
\mathcal{F}_{\sigma}(\bm{k}),
\label{del_0}
\\
\Delta_{s}^{c} &= -\dfrac{V}{N} \sum_{\bm{k}, \sigma} \Delta(\bm{k})\eta_{s}(\bm{k})
\mathcal{F}_{\sigma}(\bm{k}),
\label{del_s}
\\
\Delta_{d}^{c} &= -\dfrac{V}{N} \sum_{\bm{k}, \sigma} \Delta(\bm{k})\eta_{d}(\bm{k})
\mathcal{F}_{\sigma}(\bm{k}),
\label{del_d}
\end{align}
where $N$ denotes the number of sites, $\sigma=\pm1$, $\eta_{s}({\bm{k}}) = (\cos k_{x} + \cos k_{y})/2$, $\eta_{d}({\bm{k}}) = (\cos k_{x} - \cos k_{y})/2$
and 
\begin{align}
\mathcal{F}_{\sigma}(\bm{k})= \frac{1}{E^{\sigma}_{\bm{k}}}\tanh(\frac{\beta E^{\sigma}_{\bm{k}}}{2}).
\label{F_k}
\end{align}
Here $\beta$ is the inverse of temperature and
\begin{equation}
E^{\sigma}_{\bm{k}} = \sqrt{\varepsilon_{\sigma}^2(\bm{k}) + \vert \Delta(\bm{k}) \vert^2 }
\label{eign_2}
\end{equation}
are the two excitation spectra of the BdG Hamiltonian, where $\varepsilon_{\sigma}(\bm{k}) = \xi(\bm{k}) + \sigma l(\bm{k})$ refers to the normal-state spectra with $l(\bm{k})$ the magnitude of the $\bm{l}(\bm{k})$ vector.

To capture the phases of the three pairings, we define 
$\Delta_{\alpha}^{c} = \Delta_\alpha e^{i\phi_\alpha}$ for $\alpha \in \{0,s,d \}$,
with $\Delta_{\alpha}$ and $\phi_{\alpha}$ being real numbers.
Accordingly, the three complex self-consistent equations 
given by Eqs.~(\ref{del_0})-(\ref{del_d})
can be separated into six real equations. By solving the self-consistent 
equations numerically, we find that, when both
$s$- and $d$-wave superconducting order parameter are nonvanishing, their phases favor $\phi_s = \phi_0$ and $\phi_d = \phi_0 \pm \pi/2$ (``$\pm$'' are degenerate in energy). Therefore, the superconducting order parameter can be written explicitly as
\begin{align}
\Delta(\bm{k}) = \Delta_0 + \Delta_s\eta_{s}({\bm{k}}) + i \Delta_d\eta_{d}({\bm{k}}),
\label{dell}
\end{align}
and, accordingly, the six real self-consistent superconducting gap equations are reduced to
\begin{align}
\Delta_{0} &= -\dfrac{U}{4N} \sum_{\bm{k}, \sigma}  \big\{\Delta_{0}+\Delta_{s}\eta_{s}(\bm{k}) \big\} \mathcal{F}_{\sigma}(\bm{k}),
\\
\Delta_{s} &= -\dfrac{V}{N} \sum_{\bm{k}, \sigma} \eta_{s}(\bm{k}) \big\{\Delta_{0}+\Delta_{s}\eta_{s}(\bm{k}) \big\} \mathcal{F}_{\sigma}(\bm{k}),
\\
1 &= -\dfrac{V}{N} \sum_{\bm{k}, \sigma} \eta^2_{d}(\bm{k})\mathcal{F}_{\sigma}(\bm{k}).
\end{align}
It should be noted that, when $s$- and $d$-wave pairing coexist, the last term of the BdG Hamiltonian (\ref{Hami_1}) should be rewritten as $\big\{ [\Delta_0 +\Delta_{s}\eta_{s}(\bm{k}) ] \tau_x
- \Delta_{d}\eta_{d}(\bm{k}) \tau_y \big\} s_0$.
Throughout this Letter, we set the hopping amplitude $t=1$ as the energy unit and $\{\lambda_{R}, U, V\}=\{0.3, 2, -5\}$, unless we clearly mention otherwise. However, they are not unique, and different sets of parameters will yield a qualitatively similar phase diagram.

\begin{figure}[t!]
\centering
\includegraphics[scale=0.46]{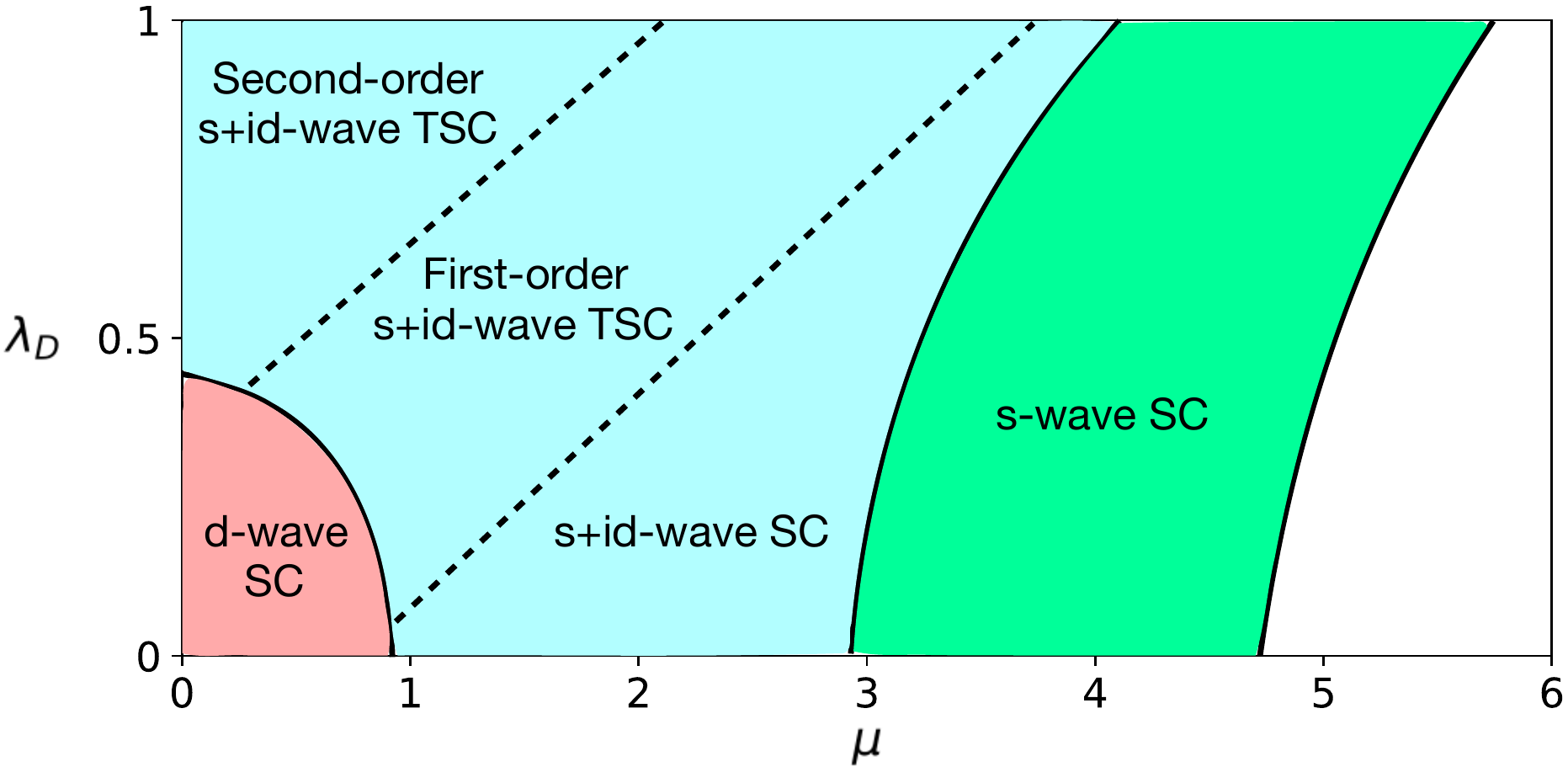}
\caption{(Color online) Zero-temperature phase diagram for $\{t, \lambda_{R}, U, V\}=\{1, 0.3, 2, -5\}$. The phase diagram contains $d$-wave SC (red color region), $s+id$-wave SC (blue color region), and $s$-wave SC (green color region). The time-reversal symmetry breaking $s+id$-wave SC phase consists of three topologically distinct phases, including first- and second-order TSC and topologically trivial SC.}
\label{fig_a}
\end{figure}

{\it Results.---}We first perform the self-consistent calculations at zero temperature. For definiteness, we consider that only $\lambda_{D}$ and $\mu$ are tunable parameters. We restrict ourselves to the positive parameter regime and present the corresponding phase diagram in Fig.~\ref{fig_a}. 
The result reveals the existence of three distinct types of pairing including time-reversal symmetry (TRS) preserving $s$- and $d$-wave, as well as the TRS breaking $s+id$-wave. The $s$- and $s+id$-wave pairing regimes are gapped except at some critical lines and points, respectively \cite{n2}, while the $d$-wave pairing regime corresponds to a nodal superconducting phase since it cannot open a bulk gap. Moreover, the TRS preserving gapped $s$-wave pairing belongs to the symmetry class DIII characterized by a $\mathbb{Z}_{2}$ invariant $\nu$ \cite{PhysRevB.81.134508,n3} and the TRS preserving gapless $d$-wave pairing
is considered as a topologically nontrivial phase, though gapless, because it harbors topologically protected gapless Majorana modes on the boundary.

Let us now focus on the interesting regime with $s+id$-wave pairing which consists of three TRS breaking topologically distinct phases, including first- and second-order TSC and topologically trivial SC, as shown in the blue color region of Fig.~\ref{fig_a}.
Since the TRS is broken in this regime, the system belongs to the symmetry class D characterized by the Chern number \cite{RevModPhys.88.035005}. To reveal the underlying topological property in a simple and transparent way, here we perform a basis transformation which maps the combination of SOC and even-parity $s+id$-wave pairing to an effective odd-parity pairing \cite{PhysRevB.81.125318}.
After the transformation (see Supplemental Material \cite{supplemental}), 
the four-band BdG Hamiltonian can be decoupled into two independent parts, 
{\it i.e.}, $\mathcal{H}(\bm{k})=\mathcal{H}_{+}(\bm{k})\oplus\mathcal{H}_{-}(\bm{k})$, with 
\begin{eqnarray}
\mathcal{H}_{\pm}(\bm{k})=\left(
                            \begin{array}{cc}
                              |\xi_{k}|\pm l(\bm{k}) & \Delta_{\pm}(\bm{k}) \\
                              \Delta_{\pm}^{*}(\bm{k}) & -|\xi_{k}| \mp l(\bm{k}) \\
                            \end{array}
                          \right),
\label{eq_no_hz}
\end{eqnarray} 
where $\Delta_{\pm}(\bm{k})=(\Delta_0 + \Delta_s\eta_{s}({\bm{k}})\mp  i \Delta_d\eta_{d}({\bm{k}}))(l_{x}\mp i l_{y})/l(\bm{k})$. 
It is apparent that $\Delta_{\pm}(-\bm{k})=-\Delta_{\pm}(\bm{k})$, confirming 
the odd-parity nature. As is known, the band topology of an odd-parity superconductor is determined by the relative configuration of the FSs and the pairing nodes, and there exists a simple relation between the Chern number ($C$) and the number of FSs ($N_{F}$) enclosing one time-reversal invariant point, which is $(-1)^{C}=(-1)^{N_{F}}$ \cite{sato2010topological,n4}. The number of FSs of $\mathcal{H}(\bm{k})$ must be even since the normal state has TRS, which implies that $C$ must be an even integer. In addition, the gapped energy spectra of $\mathcal{H}_{+}(\bm{k})$ implies the absence of a FS, which is defined as the constant-energy contour satisfying $|\xi_{k}|+l(\bm{k})=0$, while $\mathcal{H}_{-}(\bm{k})$ has either zero or two FSs, depending on the chemical potential $\mu$. As the absence of FSs always implies a trivial superconductor, only the situation that  $\mathcal{H}_{-}(\bm{k})$ has two FSs is of interest. When $C$ is a nonzero even integer, the system corresponds to a first-order TSC with $C$ branches of Majorana chiral edge states. However, when $C=0$, the system is either a topologically trivial superconductor or a second-order TSC, depending on whether 
the two FSs of $\mathcal{H}_{-}(\bm{k})$ can be continuously deformed to annihilate with each other without crossing any removable Dirac pairing nodes (not at time-reversal invariant points) or not.

Interestingly, we notice that $\mathcal{H}_{-}(\bm{k})$ takes a form similar to the toy model realizing second-order TSC proposed in Ref.~\cite{PhysRevLett.123.177001}. Here, there are four removable Dirac pairing nodes whose net sum of  winding number [defined as $\omega=(1/2\pi i)\oint\Delta_{-}^{-1}\partial_{k}\Delta_{-} dk$, with the closed integration contour enclosing only the interested pairing node] is zero lying between the two FSs, the system realizes a second-order TSC. Another way to understand 
this picture is via the edge theory. To be specific, when the 
four removable Dirac pairing nodes of $\mathcal{H}_{-}(\bm{k})$ lie between the two FSs, it means that, if we neglect the $d$-wave pairing, the line nodes of 
$s$-wave pairing (satisfying $\Delta_{0}+\Delta_{s}\eta_{s}=0$) can be chosen to lie between the two FSs. 
Since without the $d$-wave pairing, the full Hamiltonian restores the TRS, then according to the formula $\nu=\prod_{i}[sgn(\Delta_{i})]^{m_{i}}$\cite{PhysRevB.81.134508} we have $\nu=-1$, 
indicating the realization of a first-order time-reversal invariant TSC that 
hosts a pair of helical Majorana edge states. Bringing back 
the $d$-wave pairing, the helical edge states are gapped out due to the breaking of TRS. However, as 
the $d$-wave pairing itself has line nodes along the directions 
$k_{x}=\pm k_{y}$, four Majorana zero modes will be left 
at the four corners when we use open-boundary conditions in both $x$ and $y$ directions \cite{wang2018weak,wu2019higher}. 

\begin{figure*}[th!]
\centering
\includegraphics[scale=0.515]{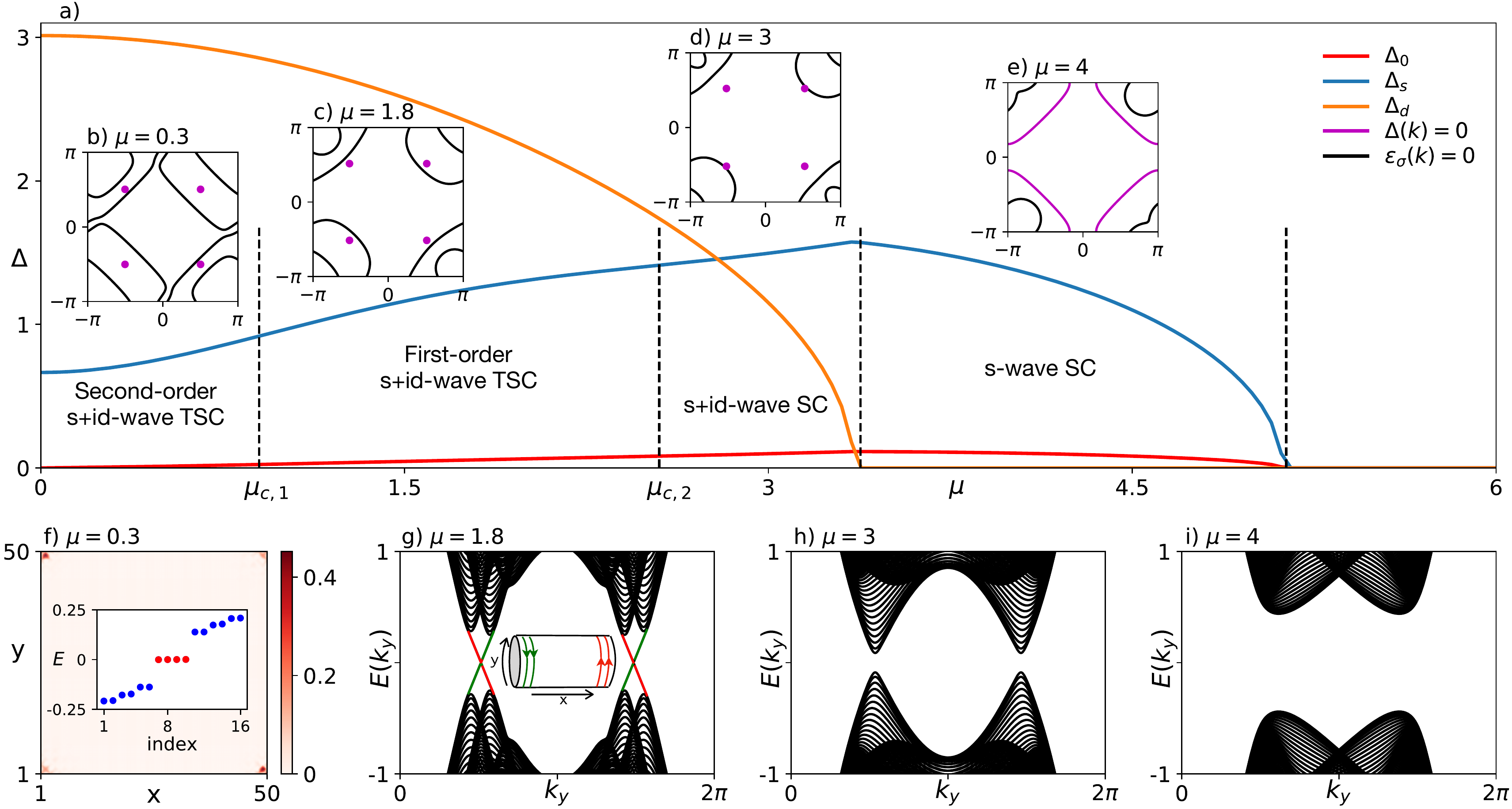}
\caption{(Color online) Common parameters: $\{t,\lambda_{R}, \lambda_{D}, U, V\}=\{1, 0.3, 0.6, 2, -5\}$ (a) The dependence of $\Delta_{0,s,d}$ on $\mu$ at zero temperature. The  dashed lines perpendicular to the $\mu$ axis correspond to phase boundaries. The four insets (b)--(e) show four representative 
configurations of FSs (black) and pairing nodes (purple points or lines) in the four distinct phases. (b) $\mu=0.3$, (c) $\mu=1.8$, (d) $\mu=3$, and (e) $\mu=4$ [for this value, one of the FS becomes a point at $(\pi,\pi)$]. (f) The probability density profiles of four Majorana corner states (red points of the inset) and the eigenvalues of the BdG Hamiltonian around zero energy in real space for a $50 \times 50$ square lattice with open-boundary conditions in both $x$ and $y$ directions. (g)--(i) Energy spectrum for cylindrical geometry with open-boundary condition only along the $x$ direction. The inset of (g) shows the distribution of chiral Majorana edge states in real space.}
\label{fig_b}
\end{figure*}

Based on the above analysis,  we find that,
within the $s+id$-wave pairing regime, the change of topology only takes place when the FSs cross the removable Dirac pairing nodes at which both $\Delta_{0}+\Delta_{s}\eta_{s}=0$ and $\Delta_{d}\eta_{d}=0$ are simultaneously fulfilled. As for 
the parameters considered, we find $\Delta_{0} \ll  \Delta_{s}, \Delta_{d}$; these nodes are almost fixed at the four points
$\bm{Q}_{\pm,\pm}=(\pm \pi/2, \pm \pi/2)$. Therefore,
the condition for topological phase transitions can be very accurately described by the normal-state condition $|\xi_{\bm{Q}_{\pm,\pm}}|-l(\bm{Q}_{\pm,\pm})=0$. It is straightforward to find that the solutions give two straight lines satisfying $|\mu|-2\sqrt{2}|\lambda_{R}\pm \lambda_{D}|=0$ (see Supplemental Material \cite{supplemental}), which correspond to the two dashed lines in the blue color region of Fig.~\ref{fig_a}.

To support the above analysis, we further diagonalize the mean-field BdG Hamiltonian in real space (see Supplemental Material \cite{supplemental}). To be specific, we fix $\lambda_{D}=0.6$ and study the evolution 
of boundary modes with $\mu$. The results are presented in Fig.~\ref{fig_b}. In accordance with the phase diagram 
in Fig.~\ref{fig_a}, we know that $\mu_{c,1}\simeq3\sqrt{2}/5\simeq0.85$ and $\mu_{c,2}\simeq9\sqrt{2}/5\simeq2.55$
are two critical points in the regime with $s+id$-wave pairing \cite{supplemental}.

\begin{figure}[!b]
\centering
\includegraphics[scale=0.49]{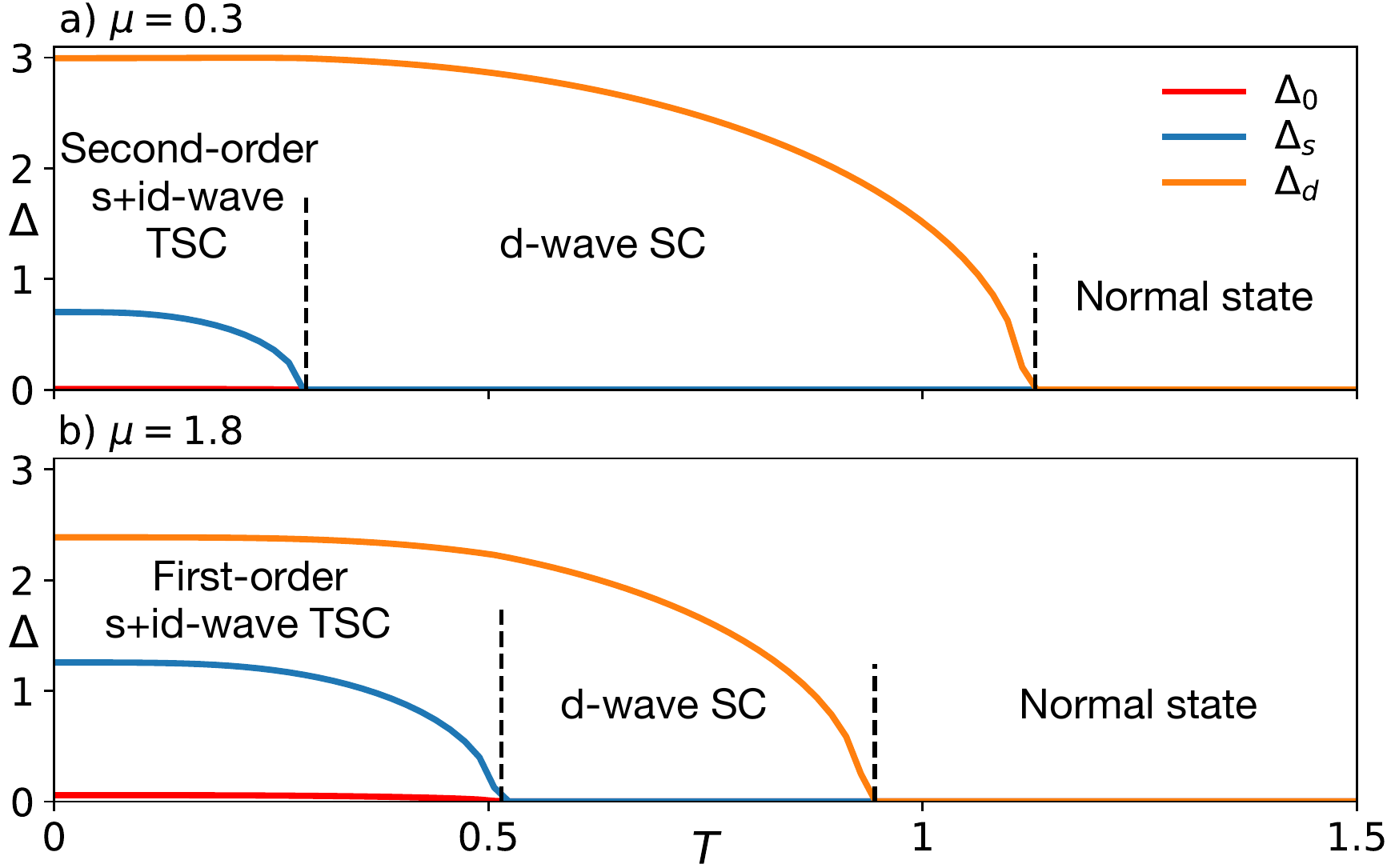}
\caption{(Color online) Temperature-driven topological phase transitions. $\{t,\lambda_{R},\lambda_{D}, U, V\}=\{1,0.3, 0.6, 2, -5\}$, and (a) $\mu = 0.3$, (b) $\mu = 1.8$. Tuning the temperature 
can change both the favored pairing and the underlying topology.}
\label{fig_c}
\end{figure}

Within each phase, we show one representative configuration of 
FSs and pairing nodes [Figs.~\ref{fig_b}(b)--\ref{fig_b}(e)]. Figure.~\ref{fig_b}(b) shows that, within the regime $0<\mu<\mu_{c,1}$, the four pairing nodes at $(\pm \pi/2, \pm \pi/2)$ are located between the two concentric FSs 
enclosing $(\pi,\pi)$, indicating the realization of a second-order TSC \cite{PhysRevLett.123.177001}. To demonstrate
this phase, we consider a square sample 
with open-boundary conditions in both $x$ and $y$ directions. A diagonalization of
the real-space Hamiltonian does confirm the existence of four Majorana corner modes
[see Fig.~\ref{fig_b}(f)] and, therefore, the realization of a second-order TSC. Within the regime 
$\mu_{c,1}<\mu<\mu_{c,2}$, only two of the four pairing nodes at $(\pm \pi/2, \pm \pi/2)$
remain to be located between the two FSs, as shown in Fig.~\ref{fig_b}(c). 
As one pairing node takes the same winding number as its inversion partner, the transition from the configuration in Fig.~\ref{fig_b}(b) to that in Fig.~\ref{fig_b}(c)
suggests a change of Chern number by two. In other words, a first-order TSC with $C=2$ is realized in the regime $\mu_{c,1}<\mu<\mu_{c,2}$. 
To demonstrate this phase, we consider a cylinder geometry with open-boundary conditions only 
along the $x$ direction. The numerical result confirms
the existence of two chiral Majorana modes on each edge [see Fig.~\ref{fig_b}(g)] and, therefore, 
the realization of a first-order TSC with $C=2$. Remarkably, the above results suggest that a topological phase transition between second- and first-order TSCs takes place at $\mu_{c,1}$ \cite{n1}. Numerical calculations reveal the absence of gapless boundary modes [Figs.~\ref{fig_b}(h) and \ref{fig_b}(i)] in the regime $\mu>\mu_{c,2}$ indicating that the Hamiltonian is trivial in topology in this regime.

So far, we have restricted the results to the zero-temperature limit. By performing self-consistent calculations at finite temperature, we find that, for a given configuration of FSs, different pairing types exhibit different 
temperature dependence. As a result, the favored pairing can 
undergo a dramatic change at some critical temperature. To be specific, Fig.~\ref{fig_c} shows two examples whose ground states at zero temperature are second- and first-order $s+id$-wave TSCs. From this figure, it is readily seen that the increase of temperature leads to a change of the favored pairing from gapped $s+id$-wave TSC to $d$-wave gapless SC at a parameter-dependent critical temperature. Since the $d$-wave  pairing leads to the realization of nodal or Dirac SC, it indicates that the temperature itself provides a way to tune 
the underlying topological properties.

{\it Conclusions.---}In this Letter, we showed that both first- and second-order TRS breaking topological superconductivity as well as the topological phase transition between them can emerge in the extended Hubbard model with both Rashba and Dresselhaus SOC, even in the absence of an external magnetic field or magnetic order. Moreover, we demonstrated that, with appropriate FS structure, tuning only the temperature can result in interesting topological phase transitions in this system. Our findings are relevant to many systems where both SOCs and interactions are tunable, including InSb \cite{PhysRevB.81.125318} or InGaAs \cite{PhysRevLett.78.1335} quantum  wells in  proximity to a high-temperature iron-based superconductor, which has an order parameter with $s+id$-wave superconducting pairing symmetry \cite{PhysRevLett.102.217002, RevModPhys.83.1589, PhysRevLett.121.076401}, oxide interfaces like LaAlO$_{3}$/SrTiO$_{3}$ \cite{reyren2007superconducting, Ben2010soc}, and cold atom systems \cite{PhysRevB.83.140510}.

This work was supported in part by the Natural Sciences and Engineering Research Council of Canada (NSERC). Z.Y. is supported by Startup Grant of Sun Yat-sen University (No.~74130-18841219) and NSFC-11904417. Y.~N. was partially supported by JSPS KAKENHI Grant Number 18K11345 and 18K03552, the ``Topological Materials Science'' (No.~JP18H04228) KAKENHI on Innovative Areas from JSPS of Japan.

\bibliography{ref.bib}

\begin{thebibliography}{113}
\bibitem [{n2()}]{n2}%
  \BibitemOpen
  \href@noop {} {\ }\bibinfo {note} {Here, by pure $s$-wave pairing we mean
  $\Delta_0 + \Delta_s \eta_s(\bm{k})$ when at least one of the $\Delta_0$ or
  $\Delta_s$ values is finite. It should be noted that pure $s$-wave pairing
  has pairing line nodes only when $|\Delta_0/\Delta_s| \leqslant
  1$.}\BibitemShut {Stop}%
\bibitem [{n3()}]{n3}%
  \BibitemOpen
  \href@noop {} {\ }\bibinfo {note} {Here, $\mathbb{Z}_{2}$ invariant is
  defined as $\nu=\prod_{i}[sgn(\Delta_{i})]^{m_{i}}$ where $sgn(\Delta_{i})$
  represents the sign of the pairing on the $i$-th spin-split FS, and $m_{i}$
  denotes the number of time-reversal invariant points within the $i$-th
  FS.}\BibitemShut {Stop}%
\bibitem [{n4()}]{n4}%
  \BibitemOpen
  \href@noop {} {\ }\bibinfo {note} {Another important point is that if the
  normal state of a superconductor does not have a FS, the underlying topology
  must be trivial since such a phase can continuously transform to the limit
  $|\mu|\rightarrow\infty$ without the closing of either the bulk or the
  boundary gap.}\BibitemShut {Stop}%
\bibitem [{n1()}]{n1}%
  \BibitemOpen
  \href@noop {} {\ }\bibinfo {note} {When the Fermi surface is anisotropic, it
  is a general feature that topologically trivial and second-order topological
  $s+id$-wave phases are separated by an intervening first-order topological
  phase. This fact can be understood by inspecting the configurations of both
  Fermi surfaces and pairing nodes. It is not hard to see that when the Fermi
  surface is anisotropic, the four pairing nodes can not simultaneously move
  outside all Fermi surfaces.}\BibitemShut {Stop}%
\end{thebibliography}

\clearpage


\section*{Supplementary material}

# Supplemental Material for "First- and Second-Order Topological Superconductivity and Temperature-Driven Topological Phase Transitions in the Extended Hubbard Model with Spin-Orbit Coupling"

Majid Kheirkhah,[1] Zhongbo Yan,[2] Yuki Nagai,[3,4] and Frank Marsiglio[1]

[1]*Department of Physics, University of Alberta, Edmonton, Alberta T6G 2E1, Canada*
[2]*School of Physics, Sun Yat-Sen University, Guangzhou 510275, China*
[3]*CCSE, Japan Atomic Energy Agency, 178-4-4, Wakashiba, Kashiwa, Chiba, 277-0871, Japan*
[4]*Mathematical Science Team, RIKEN Center for Advanced Intelligence Project (AIP),
1-4-1 Nihonbashi, Chuo-ku, Tokyo 103-0027, Japan*

This supplemental material provides the details of the derivations of some important equations presented in the main text. It consists of five sections: (I) The gap equations in momentum space, (II) the Bogoliubov-de Gennes Hamiltonian in real space, (III) effective realization of odd-parity superconductivity, (IV) phase boundaries in the $s+id$-wave pairing regime, and (V) phase diagram under different parameter conditions.

## I. THE GAP EQUATIONS IN MOMENTUM SPACE

As the normal state has time-reversal symmetry and we are interested in the superconducting pairing instability, the last line of the real-space Hamiltonian in the main text can be written in momentum space as,

$$H_{int} = \frac{1}{N} \sum_{\bm{k},\bm{q}} (U + V\zeta_{\bm{k}-\bm{q}}) c^\dagger_{\bm{k},\uparrow} c^\dagger_{-\bm{k},\downarrow} c_{-\bm{q},\downarrow} c_{\bm{q},\uparrow}, \tag{S1}$$

where $\zeta_{\bm{k}-\bm{q}} = \sum_{\bm{\delta}} e^{-i(\bm{k}-\bm{q})\cdot\bm{\delta}} = 2\{\cos(k_x - q_x) + \cos(k_y - q_y)\}$ and $\bm{\delta}$ is the coordinates of the four nearest-neighbours of a given site in a square lattice with unit lattice constant. Note that we have dropped terms corresponding to pairing with non-zero center-of-mass momentum. Performing mean-field decoupling by considering only the even-parity superconducting order-parameter and ignoring constant terms leads to the following quadratic Hamiltonian,

$$H^{MF}_{int} = \frac{1}{2} \sum_{\bm{k}} \Delta_{\alpha\beta}(\bm{k}) c^\dagger_{\bm{k},\alpha} c^\dagger_{-\bm{k},\beta} + h.c., \tag{S2}$$

where the even-parity superconducting order-parameter can be written as $\Delta_{\alpha\beta}(\bm{k}) = i\Delta(\bm{k}) s_y^{\alpha\beta}$ and

$$\Delta(\bm{k}) = -\frac{1}{N} \sum_{\bm{q}} \left\{ U + \frac{V}{2}(\zeta_{\bm{k}+\bm{q}} + \zeta_{\bm{k}-\bm{q}}) \right\} \langle c_{\bm{q}\uparrow} c_{-\bm{q}\downarrow} \rangle. \tag{S3}$$

It is obvious that $\Delta(-\bm{k}) = \Delta(\bm{k})$ by using of the fact that $\zeta_{\bm{k}}$ is an even function of $\bm{k}$, i.e. $\zeta_{\bm{k}} = \zeta_{-\bm{k}}$. It can be shown that,

$$\zeta_{\bm{k}+\bm{q}} + \zeta_{\bm{k}-\bm{q}} = 8\{\eta_s(\bm{k})\eta_s(\bm{q}) + \eta_d(\bm{k})\eta_d(\bm{q})\}, \tag{S4}$$

where $\eta_s(\bm{k}) = (\cos k_x + \cos k_y)/2$, and $\eta_d(\bm{k}) = (\cos k_x - \cos k_y)/2$. This leads to the following superconducting gap functions,

$$\Delta(\bm{k}) = \Delta_0^c + \Delta_s^c \eta_s(\bm{k}) + \Delta_d^c \eta_d(\bm{k}). \tag{S5}$$

where $\Delta_{0,s,d}^c$ are some complex numbers and are given by Eq. (4) to Eq. (6) of the main paper. If we define $\Delta_\alpha^c = \Delta_\alpha e^{i\phi_\alpha}$ for $\alpha \in \{0, s, d\}$ where all the $\Delta_\alpha$ and $\phi_\alpha$ are real numbers, those three complex equations are separated into six real equations. We restrict ourselves to the even-parity $s+id$-wave pairing which is more favorable in energy by setting $\phi_s = \phi_0$ and $\phi_d - \phi_0 = \pm\pi/2$. Therefore, the resulting $s+id$-wave pairing order-parameter reads as,

$$\Delta(\bm{k}) = \Delta_0 + \Delta_s \eta_s(\bm{k}) + i\Delta_d \eta_d(\bm{k}). \tag{S6}$$

Finally, the total Hamiltonian in the Nambu space can be written as,

$$H = \frac{1}{2} \sum_{\boldsymbol{k}} \begin{pmatrix} \boldsymbol{c}_{\boldsymbol{k}}^\dagger & \boldsymbol{c}_{-\boldsymbol{k}} \end{pmatrix} \mathcal{H}(\boldsymbol{k}) \begin{pmatrix} \boldsymbol{c}_{\boldsymbol{k}} \\ \boldsymbol{c}_{-\boldsymbol{k}}^\dagger \end{pmatrix}, \tag{S7}$$

where $\boldsymbol{c}_{\boldsymbol{k}}^\dagger = (c_{\boldsymbol{k}\uparrow}^\dagger,\ c_{\boldsymbol{k}\downarrow}^\dagger)$ and $\boldsymbol{c}_{-\boldsymbol{k}} = (c_{-\boldsymbol{k}\uparrow},\ c_{-\boldsymbol{k}\downarrow})$, as well as

$$\mathcal{H}(\boldsymbol{k}) = \xi(\boldsymbol{k})\tau_z s_0 + l_x(\boldsymbol{k})\tau_0 s_x + l_y(\boldsymbol{k})\tau_z s_y - (\Delta_0 + \Delta_s \eta_{\boldsymbol{k}}^s)\tau_y s_y - \Delta_d \eta_{\boldsymbol{k}}^d \tau_x s_y, \tag{S8}$$

where $\xi(\boldsymbol{k}) = -2t(\cos k_x + \cos k_y) - \mu$, $l_x(\boldsymbol{k}) = 2(\lambda_R \sin k_y + \lambda_D \sin k_x)$, and $l_y(\boldsymbol{k}) = -2(\lambda_R \sin k_x + \lambda_D \sin k_y)$. The Pauli matrices in spin and particle-hole spaces are denoted by $s_{0,x,y,z}$ and $\tau_{0,x,y,z}$, respectively. Note that in the main paper a different Nambu basis was used.

## II. THE BOGOLIUBOV-DE GENNES HAMILTONIAN IN REAL-SPACE

The mean-field real-space Hamiltonian can be rewritten as,

$$H = \frac{1}{2} \sum_{ij} \begin{pmatrix} \boldsymbol{c}_i^\dagger & \boldsymbol{c}_i \end{pmatrix} \begin{pmatrix} A_{ij} & B_{ij} \\ -B_{ij}^* & -A_{ij}^* \end{pmatrix} \begin{pmatrix} \boldsymbol{c}_j \\ \boldsymbol{c}_j^\dagger \end{pmatrix}, \tag{S9}$$

where $\boldsymbol{c}_i^\dagger = (c_{i,\uparrow}^\dagger,\ c_{i,\downarrow}^\dagger)$ and $\boldsymbol{c}_i = (c_{i,\uparrow},\ c_{i,\downarrow})$ for $i = (i_x, i_y)$. The hermitian property of $H$, and the canonical anti-commutation relation $\{c_i, c_j^\dagger\} = \delta_{ij}$ imply that the $2N \times 2N$ complex matrix $A$ and $B$ satisfy $A^\dagger = A$ and $B^\mathrm{T} = -B$ relations, respectively, where $N$ is the number of sites and T stands for transpose. The components of $A$ and $B$ matrices are given by,

$$A_{ij} = s_0 A_{ij}^0 + s_x A_{ij}^x + s_y A_{ij}^y, \tag{S10}$$
$$B_{ij} = s_y(\mathrm{i}\Delta_{ij}^s - \Delta_{ij}^d), \tag{S11}$$

where

$$A_{ij}^0 = -t(\delta_{j,i+\hat{x}} + \delta_{j,i-\hat{x}} + \delta_{j,i+\hat{y}} + \delta_{j,i-\hat{y}}) - \mu\delta_{ij}, \tag{S12}$$
$$A_{ij}^x = -\mathrm{i}\lambda_R(\delta_{j,i+\hat{y}} - \delta_{j,i-\hat{y}}) - \mathrm{i}\lambda_D(\delta_{j,i+\hat{x}} - \delta_{j,i-\hat{x}}), \tag{S13}$$
$$A_{ij}^y = +\mathrm{i}\lambda_R(\delta_{j,i+\hat{x}} - \delta_{j,i-\hat{x}}) + \mathrm{i}\lambda_D(\delta_{j,i+\hat{y}} - \delta_{j,i-\hat{y}}), \tag{S14}$$
$$\Delta_{ij}^s = \Delta_0 \delta_{ij} + \frac{\Delta_s}{4}(\delta_{j,i+\hat{x}} + \delta_{j,i-\hat{x}} + \delta_{j,i+\hat{y}} + \delta_{j,i-\hat{y}}), \tag{S15}$$
$$\Delta_{ij}^d = \frac{\Delta_d}{4}(\delta_{j,i+\hat{x}} + \delta_{j,i-\hat{x}} - \delta_{j,i+\hat{y}} - \delta_{j,i-\hat{y}}). \tag{S16}$$

In this section and the main paper, the symbol i denoting pure imaginary number should not be confused with the site-index that generally occurs subscripted.

## III. EFFECTIVE REALIZATION OF ODD-PARITY SUPERCONDUCTIVITY

For generality, here we further include an out-of-plane Zeeman field. Accordingly, the normal state Hamiltonian is

$$H_0 = \sum_{\boldsymbol{k}} (c_{\boldsymbol{k},\uparrow}^\dagger,\ c_{\boldsymbol{k},\downarrow}^\dagger) \begin{pmatrix} \xi_{\boldsymbol{k}} + h_z & l_x - il_y \\ l_x + il_y & \xi_{\boldsymbol{k}} - h_z \end{pmatrix} \begin{pmatrix} c_{\boldsymbol{k},\uparrow} \\ c_{\boldsymbol{k},\downarrow} \end{pmatrix}, \tag{S17}$$

The spin singlet-pairing superconductivity is described by

$$H_{\mathrm{SC}} = \sum_{\boldsymbol{k}} (\Delta_s(\boldsymbol{k}) + i\Delta_d(\boldsymbol{k})) c_{\boldsymbol{k},\uparrow}^\dagger c_{-\boldsymbol{k},\downarrow}^\dagger + h.c., \tag{S18}$$

where $\Delta_s(\boldsymbol{k}) = \Delta_0 + \Delta_s\eta_s(\boldsymbol{k})$ and $\Delta_d(\boldsymbol{k}) = \Delta_d\eta_d(\boldsymbol{k})$. We first follow Ref. [S1] and do the following transformation

$$c_{\boldsymbol{k},\uparrow} = (\cos\frac{\theta_{\boldsymbol{k}}}{2} c_{\boldsymbol{k},+} + e^{-i\phi_{\boldsymbol{k}}}\sin\frac{\theta_{\boldsymbol{k}}}{2} c_{\boldsymbol{k},-}), \tag{S19}$$

$$c_{\boldsymbol{k},\downarrow} = (e^{i\phi_{\boldsymbol{k}}}\sin\frac{\theta_{\boldsymbol{k}}}{2} c_{\boldsymbol{k},+} - \cos\frac{\theta_{\boldsymbol{k}}}{2} c_{\boldsymbol{k},-}), \tag{S20}$$

where $\theta_{\boldsymbol{k}}$ satisfies $\cos\theta_{\boldsymbol{k}} = h_z/\Lambda_{\boldsymbol{k}}$ with $\Lambda_{\boldsymbol{k}} = \sqrt{h_z^2 + l_x^2 + l_y^2}$, and $\phi_{\boldsymbol{k}}$ satisfy $e^{i\phi_{\boldsymbol{k}}} = (l_x + il_y)/\sqrt{l_x^2 + l_y^2}$. Substituting Eq. (S19) and Eq. (S20) into Eq. (S17), we have

$$H_0 = \sum_{\boldsymbol{k}} (\xi_{\boldsymbol{k}} + \Lambda_{\boldsymbol{k}}) c_{\boldsymbol{k},+}^\dagger c_{\boldsymbol{k},+} + (\xi_{\boldsymbol{k}} - \Lambda_{\boldsymbol{k}}) c_{\boldsymbol{k},-}^\dagger c_{\boldsymbol{k},-}. \tag{S21}$$

For the superconducting part,

$$H_{\text{SC}} = \sum_{\boldsymbol{k}} \left( -\frac{h_z(\Delta_s(\boldsymbol{k}) + i\Delta_d(\boldsymbol{k}))}{\Lambda_{\boldsymbol{k}}} c_{\boldsymbol{k},+}^\dagger c_{-\boldsymbol{k},-}^\dagger + h.c. \right)$$

$$+ \frac{1}{2}\sum_{\boldsymbol{k}} \left( \frac{-(\Delta_s(\boldsymbol{k}) + i\Delta_d(\boldsymbol{k}))(l_x - il_y)}{\Lambda_{\boldsymbol{k}}} c_{\boldsymbol{k},+}^\dagger c_{-\boldsymbol{k},+}^\dagger + h.c. \right)$$

$$+ \frac{1}{2}\sum_{\boldsymbol{k}} \left( \frac{-(\Delta_s(\boldsymbol{k}) + i\Delta_d(\boldsymbol{k}))(l_x + il_y)}{\Lambda_{\boldsymbol{k}}} c_{\boldsymbol{k},-}^\dagger c_{-\boldsymbol{k},-}^\dagger + h.c. \right) \tag{S22}$$

Below we go beyond Ref. [S1] and do a band-dependent gauge transformation, $c_{\boldsymbol{k},+} = -\bar{c}_{\boldsymbol{k},+} e^{i\varphi_{\boldsymbol{k}}}$, $c_{\boldsymbol{k},-} = \bar{c}_{\boldsymbol{k},-}$, with $e^{i\varphi_{\boldsymbol{k}}} = \frac{(\Delta_s(\boldsymbol{k}) + i\Delta_d(\boldsymbol{k}))}{\sqrt{\Delta_s^2(\boldsymbol{k}) + \Delta_d^2(\boldsymbol{k})}}$. Accordingly, we have

$$H_0 = \sum_{\boldsymbol{k}} (\xi_{\boldsymbol{k}} + \Lambda_{\boldsymbol{k}}) \bar{c}_{\boldsymbol{k},+}^\dagger \bar{c}_{\boldsymbol{k},+} + (\xi_{\boldsymbol{k}} - \Lambda_{\boldsymbol{k}}) \bar{c}_{\boldsymbol{k},-}^\dagger \bar{c}_{\boldsymbol{k},-}, \tag{S23}$$

$$H_{\text{SC}} = \sum_{\boldsymbol{k}} \left( \frac{h_z \sqrt{\Delta_s^2(\boldsymbol{k}) + \Delta_d^2(\boldsymbol{k})}}{\Lambda_{\boldsymbol{k}}} \bar{c}_{\boldsymbol{k},+}^\dagger \bar{c}_{-\boldsymbol{k},-}^\dagger + h.c. \right)$$

$$+ \frac{1}{2}\sum_{\boldsymbol{k}} \left( \frac{-(\Delta_s(\boldsymbol{k}) - i\Delta_d(\boldsymbol{k}))(l_x - il_y)}{\Lambda_{\boldsymbol{k}}} \bar{c}_{\boldsymbol{k},+}^\dagger \bar{c}_{-\boldsymbol{k},+}^\dagger + h.c. \right)$$

$$+ \frac{1}{2}\sum_{\boldsymbol{k}} \left( \frac{-(\Delta_s(\boldsymbol{k}) + i\Delta_d(\boldsymbol{k}))(l_x + il_y)}{\Lambda_{\boldsymbol{k}}} \bar{c}_{\boldsymbol{k},-}^\dagger \bar{c}_{-\boldsymbol{k},-}^\dagger + h.c. \right). \tag{S24}$$

By further doing the following transformation

$$\bar{c}_{\boldsymbol{k},+} = \cos\frac{\gamma_{\boldsymbol{k}}}{2} \tilde{c}_{\boldsymbol{k},+} + \sin\frac{\gamma_{\boldsymbol{k}}}{2} \tilde{c}_{-\boldsymbol{k},-}^\dagger, \tag{S25}$$

$$\bar{c}_{-\boldsymbol{k},-}^\dagger = \sin\frac{\gamma_{\boldsymbol{k}}}{2} \tilde{c}_{\boldsymbol{k},+} - \cos\frac{\gamma_{\boldsymbol{k}}}{2} \tilde{c}_{-\boldsymbol{k},-}^\dagger, \tag{S26}$$

where $\gamma_{\boldsymbol{k}} = \arctan\frac{h_z\sqrt{\Delta_s^2(\boldsymbol{k}) + \Delta_d^2(\boldsymbol{k})}}{\Lambda_{\boldsymbol{k}}\xi_{\boldsymbol{k}}}$, the Hamiltonian can be further transformed into the below form

$$H = \sum_{\boldsymbol{k}\alpha} \epsilon_{\boldsymbol{k},\alpha} \tilde{c}_{\boldsymbol{k},\alpha}^\dagger \tilde{c}_{\boldsymbol{k},\alpha} - \frac{1}{2}\left( \frac{(\Delta_s(\boldsymbol{k}) - i\alpha\Delta_d(\boldsymbol{k}))(l_x - i\alpha l_y)}{\Lambda_{\boldsymbol{k}}} \tilde{c}_{\boldsymbol{k},\alpha}^\dagger \tilde{c}_{-\boldsymbol{k},\alpha}^\dagger + h.c. \right), \tag{S27}$$

where $\epsilon_{\boldsymbol{k},\pm} = \sqrt{\xi_{\boldsymbol{k}}^2 + \frac{(\Delta_s^2(\boldsymbol{k}) + \Delta_d^2(\boldsymbol{k}))h_z^2}{\Lambda_{\boldsymbol{k}}^2}} \pm \Lambda_{\boldsymbol{k}}$. It is readily seen that the Hamiltonian can be decoupled into two independent parts, i.e., $H = H_+ \oplus H_-$, with

$$H_+ = \sum_{\boldsymbol{k}} \epsilon_{\boldsymbol{k},+} \tilde{c}_{\boldsymbol{k},+}^\dagger \tilde{c}_{\boldsymbol{k},+} - \frac{1}{2}\left( \frac{(\Delta_s(\boldsymbol{k}) - i\Delta_d(\boldsymbol{k}))(l_x - il_y)}{\Lambda_{\boldsymbol{k}}} \tilde{c}_{\boldsymbol{k},+}^\dagger \tilde{c}_{-\boldsymbol{k},+}^\dagger + h.c. \right), \tag{S28}$$

$$H_- = \sum_{\boldsymbol{k}} \epsilon_{\boldsymbol{k},-} \tilde{c}_{\boldsymbol{k},-}^\dagger \tilde{c}_{\boldsymbol{k},-} - \frac{1}{2}\left( \frac{(\Delta_s(\boldsymbol{k}) + i\Delta_d(\boldsymbol{k}))(l_x + il_y)}{\Lambda_{\boldsymbol{k}}} \tilde{c}_{\boldsymbol{k},-}^\dagger \tilde{c}_{-\boldsymbol{k},-}^\dagger + h.c. \right). \tag{S29}$$

It is obvious that both $H_+$ and $H_-$ describe odd-parity superconductors. It is worthy pointing out the pairing forms in $H_+$ and $H_-$ are similar to that in the toy model for second-order topological odd-party superconductors proposed in Ref. [S2]. In the limit $h_z = 0$, we obtain Eq. (13) in the main paper.

## IV. PHASE BOUNDARIES IN THE $s+id$-WAVE PAIRING REGIME

As shown in the Fig. 1 of the main paper, the $s+id$-wave superconducting regime consists of three time-reversal symmetry breaking topologically distinct phases. The boundaries between these phases are two straight lines with the same slope but different intercepts. In this section, we are going to derive the equations of those two lines.

The nodes of $s+id$-wave pairing are located where both $\Delta_0 + \Delta_s \eta_s(\mathbf{k}) = 0$ and $\eta_d(\mathbf{k}) = 0$ are simultaneously satisfied in the first Brillouin zone (FBZ) while the Fermi surfaces of the normal state are determined by the equation $\varepsilon_\sigma(\mathbf{k}) = \xi(\mathbf{k}) + \sigma l(\mathbf{k}) = 0$. As explained in the main paper, a topological phase transition happens in the $s+id$-wave regime when the Fermi surface cross these pairing nodes. It can be shown that the four pairing nodes in the FBZ satisfy,

$$\frac{4\Delta_0 t}{\Delta_s} + \sigma l(\mathbf{k}) = \mu, \tag{S30}$$

or

$$\frac{4\Delta_0 t}{\Delta_s} \pm 2\sqrt{(\lambda_R \sin k_y + \lambda_D \sin k_x)^2 + (\lambda_R \sin k_x + \lambda_D \sin k_y)^2} = \mu. \tag{S31}$$

Since $\eta_d(\mathbf{k}) = 0$ are satisfied on the two lines $k_x = \pm k_y$, using the fact that $\sin k_x = \pm \sin k_y$ on these two lines leads to

$$\sin^2 k_x (\lambda_R \pm \lambda_D)^2 = \frac{1}{2}(\frac{\mu}{2} - \frac{2\Delta_0 t}{\Delta_s})^2. \tag{S32}$$

At the pairing nodes, we also have $\cos k_x = -\frac{\Delta_0}{\Delta_s}$, so,

$$\{1 - (\frac{\Delta_0}{\Delta_s})^2\}(\lambda_R \pm \lambda_D)^2 = \frac{1}{2}(\frac{\mu}{2} - \frac{2\Delta_0 t}{\Delta_s})^2. \tag{S33}$$

For $\Delta_0 \ll \Delta_s$, we have

$$(\lambda_R \pm \lambda_D)^2 \simeq \frac{\mu^2}{8}. \tag{S34}$$

If we restrict ourselves to the lines with positive slope in $\lambda_D - \mu$ plane, we will have

$$\lambda_D \simeq \frac{\sqrt{2}}{4}\mu \pm \lambda_R, \tag{S35}$$

which are the equations of two lines with the slope $\sqrt{2}/4$ and intercept $\lambda_R$ (the boundary between first and second-order TSC phases) and intercept $-\lambda_R$ (the boundary between first-order TSC and the topologically trivial phase).

## V. PHASE DIAGRAM UNDER DIFFERENT PARAMETER CONDITIONS

Although we used some specific parameters in the main paper to show some numerical results, it does not mean that those parameters are unique. In fact, different parameters can be chosen and these will yield a qualitatively similar phase diagram. Here, for example, we use smaller values for the spin-orbit coupling magnitudes compared to the ones in the main paper and show the results in Fig. S1. The result is quite similar to the one shown in Fig. 2a in the main paper.

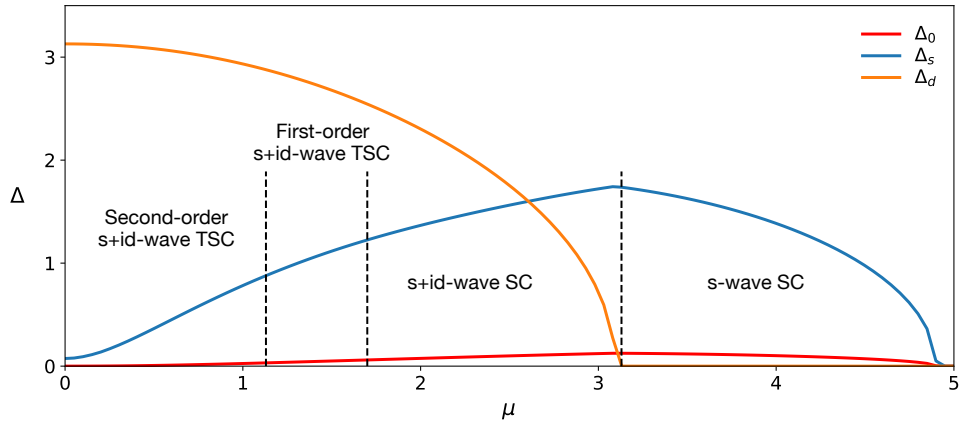

FIG. S1. (Color online) Chemical potential dependency of $\Delta_{0,s,d}$ at zero-temperature for $\{t, \lambda_R, \lambda_D, U, V\} = \{1, 0.1, 0.5, 2, -5\}$.

[S1] Jason Alicea, "Majorana fermions in a tunable semiconductor device," Phys. Rev. B **81**, 125318 (2010).
[S2] Zhongbo Yan, "Higher-order topological odd-parity superconductors," Phys. Rev. Lett. **123**, 177001 (2019).



\end{document}